\documentclass[preprint,5p,twocolumn,number,11pt]{elsarticle}

\usepackage{graphics}

\usepackage{amssymb}
\usepackage{amsthm}

\usepackage{ulem}
\usepackage{tabularx}
\usepackage{textcomp}
\usepackage[english]{babel}
\usepackage{upgreek}

\journal{Physical Review D}

\begin{document}

\begin{frontmatter}


\title{Reconstruction of the energy and depth of maximum of cosmic-ray air-showers from LOPES radio measurements}

\author[1]{W.D.~Apel}
\author[14]{J.C.~Arteaga-Velazquez}
\author[3]{L.~B\"ahren}
\author[1]{K.~Bekk}
\author[4]{M.~Bertaina}
\author[5]{P.L.~Biermann}
\author[1,2]{J.~Bl\"umer}
\author[1]{H.~Bozdog}
\author[6]{I.M.~Brancus}
\author[4,8,17]{E.~Cantoni}
\author[4]{A.~Chiavassa}
\author[1]{K.~Daumiller}
\author[15]{V.~de~Souza}
\author[4]{F.~Di~Pierro}
\author[1]{P.~Doll}
\author[1]{R.~Engel}
\author[3,9,5]{H.~Falcke}
\author[2]{B.~Fuchs}
\author[10,18]{D.~Fuhrmann}
\author[11]{H.~Gemmeke}
\author[7]{C.~Grupen}
\author[1]{A.~Haungs}
\author[1]{D.~Heck}
\author[3]{J.R.~H\"orandel}
\author[5]{A.~Horneffer}
\author[2]{D.~Huber}
\author[1]{T.~Huege\corref{cor}}
\ead{tim.huege@kit.edu}
\author[16]{P.G.~Isar}
\author[10]{K.-H.~Kampert}
\author[2]{D.~Kang}
\author[11]{O.~Kr\"omer}
\author[3]{J.~Kuijpers}
\author[2]{K.~Link}
\author[12]{P.~{\L}uczak}
\author[2]{M.~Ludwig}
\author[1]{H.J.~Mathes}
\author[2]{M.~Melissas}
\author[8]{C.~Morello}
\author[1]{J.~Oehlschl\"ager}
\author[2]{N.~Palmieri\corref{cor}}
\ead{nunzia.palmieri@kit.edu}
\author[1]{T.~Pierog}
\author[10]{J.~Rautenberg}
\author[1]{H.~Rebel}
\author[1]{M.~Roth}
\author[11]{C.~R\"uhle}
\author[6]{A.~Saftoiu}
\author[1]{H.~Schieler}
\author[11]{A.~Schmidt}
\author[1]{F.G.~Schr\"oder}
\author[13]{O.~Sima}
\author[6]{G.~Toma}
\author[8]{G.C.~Trinchero}
\author[1]{A.~Weindl}
\author[1]{J.~Wochele}
\author[12]{J.~Zabierowski}
\author[5]{J.A.~Zensus}

\address[1]{Institut f\"ur Kernphysik, Karlsruhe Institute of Technology (KIT), Germany}
\address[{14}]{Universidad Michoacana, Morelia, Mexico}
\address[3]{Radboud University Nijmegen, Department of Astrophysics, The Netherlands}
\address[4]{Dipartimento di Fisica Generale dell' Universit\`a Torino, Italy}
\address[5]{Max-Planck-Institut f\"ur Radioastronomie Bonn, Germany}
\address[2]{Institut f\"ur Experimentelle Kernphysik, Karlsruhe Institute of Technology (KIT), Germany}
\address[6]{National Institute of Physics and Nuclear Engineering, Bucharest, Romania}
\address[8]{INAF Torino, Instituto di Fisica dello Spazio Interplanetario, Italy}
\address[{15}]{Universidad S\~ao Paulo, Inst. de F\'{\i}sica de S\~ao Carlos, Brasil}
\address[9]{ASTRON, Dwingeloo, The Netherlands}
\address[{10}]{Universit\"at Wuppertal, Fachbereich Physik, Germany}
\address[{11}]{Institut f\"ur Prozessdatenverarbeitung und Elektronik, Karlsruhe Institute of Technology (KIT), Germany}
\address[7]{Universit\"at Siegen, Fachbereich Physik, Germany}
\address[{16}]{Institute for Space Sciences, Bucharest, Romania}
\address[{12}]{National Centre for Nuclear Research, Department of Cosmic Ray Physics, {\L}\'{o}d\'{z}, Poland}
\address[{13}]{University of Bucharest, Department of Physics, Romania}
\address[{17}]{now at: Istituto Nazionale di Ricerca Metrologica, Torino, Italy }
\address[{18}]{now at: Universit\"at Duisburg-Essen, Duisburg, Germany }

\cortext[cor]{Corresponding authors:}

\begin{abstract}
LOPES is a digital radio interferometer located at Karlsruhe Institute of Technology (KIT), Germany, which measures radio emission from extensive air showers at MHz frequencies in coincidence with KASCADE-Grande. In this article, we explore a method (slope method) which leverages the slope of the measured radio lateral distribution to reconstruct crucial attributes of primary cosmic rays. First, we present an investigation of the method on the basis of pure simulations. Second, we directly apply the slope method to LOPES measurements. Applying the slope method to simulations, we obtain uncertainties on the reconstruction of energy and depth of shower maximum (X$_{\mathrm{max}}$) of 13~\% and 50~g/cm$^{2}$, respectively. Applying it to LOPES measurements, we are able to reconstruct energy and X$_{\mathrm{max}}$ of individual events with upper limits on the precision of 20-25~\% for the primary energy and 95~g/cm$^{2}$ for X$_{\mathrm{max}}$, despite strong human-made noise at the LOPES site.
\end{abstract}


\end{frontmatter}

\section{Introduction}

Hundred years have passed since the discovery of cosmic radiation, and an accurate reconstruction of both the energy and mass of primary cosmic rays at high energies still remains a compelling need in contemporary astro-particle physics.

Only recently the methods for detection of coherent MHz radiation from extensive air showers as well as the understanding of the underlying emission physics have been strongly improved, reaching important milestones \cite{Huege_icrc13_ht}.
A precise reconstruction of both the primary energy and the depth of the shower maximum (X$_{\mathrm{max}}$) of air showers is a fundamental goal for modern radio detection, which aims to become competitive and complementary to the already well-established fluorescence and Cherenkov detection techniques, limited, in contrast to the radio technique, to a low duty cycle \cite{Arqueros2008}. 

In the following, we present a method (slope method), which obtains information on the energy and the depth of maximum of air showers from features of the lateral distribution of the radio signals: On the one hand, a defined distance from the shower axis exists where the reconstruction of the primary energy is affected least by shower-to-shower fluctuations. The presence of such a characteristic distance was previously predicted with REAS2 simulations \cite{REAS2}. Here, we demonstrate on the basis of state-of-the-art CoREAS simulations \cite{Ludwig_arena12} that such a characteristic distance is still present when refractive index effects and radiation due to the variation of the number of charged particles are taken into account, and that measurements at this distance can thus be exploited for an energy determination. On the other hand, the slope of the radio lateral distribution is related to the geometrical distance between the observer and the radio source \cite{REAS2,
REAS3,MGMRc}. Therefore, information on the depth of shower maximum (X$_{\mathrm{max}}$) and consecutively on the type of primary particle initiating the shower, can be directly extracted from the radio lateral distribution measurement: For iron-initiated air showers, which start earlier and develop faster in the atmosphere compared to proton-initiated air showers, the radio source is typically further away from an observer at ground, and thus the slope of the lateral distribution is flatter in comparison.
Recently, this sensitivity of the slope of the measured radio lateral distribution to the development of air showers has been proven experimentally with LOPES measurements \cite{lopes_muonLateralPaper}. 

We strive to take a practical approach tailored to the data quality provided by the LOPES experiment. In particular, we aim to develop an analysis strategy which can exploit the potential of LOPES in light of its experimental limitations (relatively low number of antennas, limited lateral distance range covered in a single measurement, relatively high background noise) without becoming overly involved and complex. For experiments with higher data quality, the methods presented here should thus be refined and further improved.

Another independent method has been separately developed and applied on LOPES measurements to extract information on X$_{\mathrm{max}}$. This method considers the shape of the radio wave front \cite{Schroeder_thesis,Schroeder_arena12}, but lower precision in the reconstruction of X$_{\mathrm{max}}$ compared to the slope method was achieved. The slope and wavefront methods could in the future also be combined with each other and with further methods such as the determination of the pulse width, or spectral slope \cite{GrebeSpectral}, respectively, or the determination of the position of the Cherenkov ring \cite{VriesCherenkovXmax}.

CoREAS simulations \cite{Ludwig_arena12}, which include a realistic treatment of the refractive index of the atmosphere, are more sophisticated and more complete than simulations with the REAS predecessor codes. The remarkable agreement between CoREAS simulations and the measured slope of the radio lateral distribution \cite{Huege_icrc13_ht,Schroeder_2013} makes studies based on CoREAS simulations even more promising than previous analyses.

An amplitude scale mismatch between the measured (LOPES) and simulated (CoREAS) radio pulses still exists \cite{Schroeder_2013}. However, it does not affect the part of this analysis in which the slope of the radio lateral distribution, i.e., the ratio of amplitudes at different distances, is considered. It only has influence when absolute amplitude values are used, i.e., in our case on the investigation of the energy reconstruction (sec.\ \ref{sec_energy}), in which it will influence the calibration constants determined for the linear correlations between primary particle energy and radio amplitude.

Taking the analysis described in \cite{Palmieri_arena12,Palmieri_thesis} as a guideline, we enhance and further improve the slope method.
As a first step, we apply the slope method directly on pure CoREAS simulations, deriving important calibration parameters. In a second step, we apply the slope method on events measured with LOPES to investigate the reconstruction of the total primary energy and the depth the shower maximum.


\section{LOPES event selection}
\label{sec_evSelection}
\begin{figure}[t]
\centering
\includegraphics[width=0.5\textwidth]{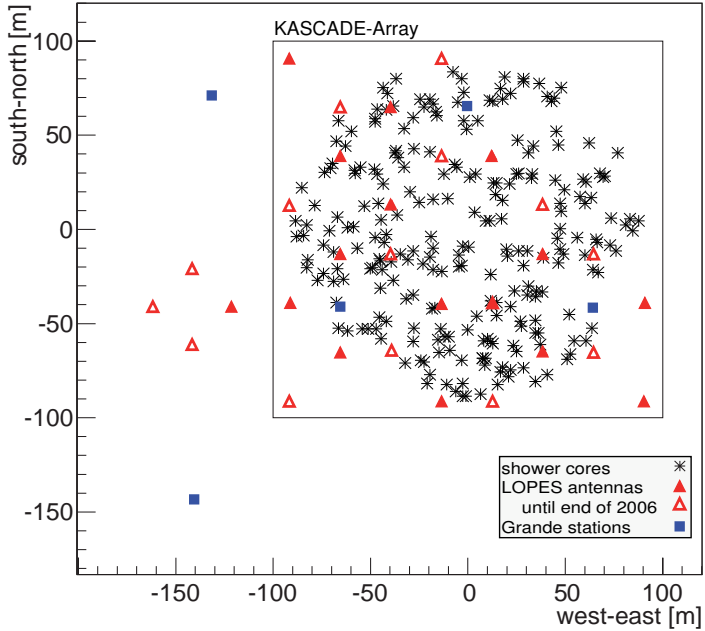}
\caption{Shower core positions of the 205 selected events, measured simultaneously by the particle detector KASCADE-Grande and by the LOPES east-west-aligned antennas.} \label{fig_map_selection}
\end{figure}

LOPES \cite{Falke_lopes,Huber_lopes3d} is one of the pioneering experiments in the digital detection of radio emission from air showers.
Placed at Karlsruhe Institute of Technology, Germany, LOPES benefits from the co-location with the particle detector experiment KASCADE-Grande \cite{kascadeG}, which provides us with the reconstruction of fundamental air-shower parameters: arrival direction, shower core, and primary energy.

Characteristic for the LOPES analysis pipeline is the interferometric combination of the recorded electric field strength traces: according to the arrival direction of the air shower, the traces detected in each individual antenna are first shifted in time. Afterwards, a cross-correlation beam (CC-beam) and a power-beam, which gives the total power received in all the antennas, are computed \cite{HornefferThesis2006,HornefferICRC2007}. The digital cross-correlation beam forming plays an important role also for the reconstruction of the lateral distribution of amplitudes. Indeed, both the CC-beam and the power-beam are crucial for the selection of events with a clear radio signal, and for the determination of the exact time of the radio pulse (CC-beam time). A Hilbert envelope is applied to the up-sampled trace of each individual antenna, and the maximum instantaneous amplitude of the radio signal is defined by the local maximum of the Hilbert envelope closest to the CC-beam time. More details can be found in 
\cite{Schroeder_thesis,Schroeder_2013,HornefferThesis2006,HornefferICRC2007}. 

The events selected for the slope method analysis presented here have been measured with the LOPES30 and LOPESpol setups \cite{Huege_arena10}. The former consisted of 30 calibrated dipole antennas, all oriented in the east-west direction, while the latter used 15 east-west and 15 north-south-aligned antennas. For the purpose of this analysis, only data of the east-west oriented channels are taken into account, primarily due to the higher available statistics \cite{HornefferICRC2007}. 
The effective frequency band used for the analysis is 43-74~MHz.

The selected events have a primary energy between 10$^{17}$ and 10$^{18}$~eV, a zenith angle less then 40$^{\circ}$ and a core position at a maximum distance of 90~m from the center of the KASCADE array - (0~m~west-east, 0~m~south-north) in fig.\ \ref{fig_map_selection}. These cuts are made to achieve high quality in the KASCADE air shower reconstruction. A high signal-to-noise ratio and high coherence for the radio signal in the antennas is required as well by forcing the signal-to-noise ratio of the CC-beam, i.e. (amplitude of the cross-correlation beam / RMS of the cross-correlation beam~$\cdot \sqrt{N_\textrm{ant}/30}$)\footnote{The CC-beam is normalized with a factor $\sqrt{N_\textrm{ant}/30}$, since not always all the 30 antennas participate in each event} to be larger than 9, and the fraction of correlated power in the antennas (CC-beam/power-beam) larger than 80~\%. Further quality cuts demand a good fit for the lateral distribution function (sec.\ \ref{sec_ldf}). All cuts are chosen to achieve high 
standards in the KASCADE-Grande data reconstruction and high quality radio measurements, yet simultaneously acquire good statistics. 205 events of almost 900 KASCADE-Grande pre-selected showers remain after applying all cuts.


\section{CoREAS simulations of the selected events}
The CoREAS code is one of the latest developments of simulation codes reflecting an increasingly precise knowledge and an improved modelling of the radio emission from air showers \cite{Ludwig_arena12}. Through the inclusion of a realistic treatment of the refractive index of the atmosphere, CoREAS simulations take into account Cherenkov-like time-compression of the radio emission occurring for observers near the Cherenkov angle. This effect has been recognized to be the main reason for the previously measured flattening of the radio lateral distribution close to the shower axis \cite{lopesLDF}. The flattening effects depend on the geometrical distance of the shower maximum to the observer, and thus on both the zenith angle of the air shower and on the type of the primary particle.
A detailed comparison between CoREAS simulations and LOPES measurements is presented in \cite{Schroeder_2013}.

For the following analysis, CORSIKA \cite{corsika} and CoREAS are used to simulate the particle cascade and the radio emission from the electromagnetic component of air showers, respectively. QGSJet~II.03 \citep{Ostapchenko_QGSjetII2006} and FLUKA \citep{FLUKA}
are used as high energy and low energy interaction models, respectively.

Specific criteria are applied to generate the CoREAS simulations to facilitate the final application of the slope method directly on LOPES data. On the one hand, the specific individual events measured by LOPES are simulated: The energy, the incoming direction and the core position of each LOPES event, reconstructed from the measurements with the particle detector KASCADE(-Grande) \cite{kascadeG}, are used as input parameters for the simulations. The simulated electric field is filtered with an ideal rectangular bandpass filter (43-74~MHz). The geomagnetic field for the simulations is set to values valid for the LOPES site (47 $\upmu$T magnetic field strength, 65$^{\circ}$ inclination angle, negligible declination). With this procedure, the geometrical acceptance and important characteristics of the LOPES experiment are taken into account. Each LOPES event is simulated as both proton and iron primary. On the other hand, the observer positions in the simulations appropriately represent the antenna positions 
in the LOPES array: 
For each individual event, the correct positions of the individual LOPES antennas with respect to the shower core are taken into account. For the selected events, the cores of the shower typically lie in between the antennas, thus the fit of the lateral distribution function (LDF) does not only involve observer positions in a specific azimuthal direction relative to the core. As a consequence, the charge excess effect which typically contributes to the total radio emission at a level of $\approx10-15$~\% \cite{REAS3,Huege_icrc13} in AERA, and can be expected to be about half as important in LOPES due to the stronger local magnetic field, is included in the following analysis. We can thus judge the robustness of our method in light of the complex asymmetries present in the actual measured radio lateral distributions.
No pre-selection of typical showers (showers with a typical X$_{\mathrm{max}}$ value) is applied to the CoREAS simulations. Therefore, shower-to-shower fluctuations are included in the investigation for the data set as a whole.

\begin{figure}[t]
\centering
\includegraphics[width=0.5\textwidth, height=0.35\textwidth]{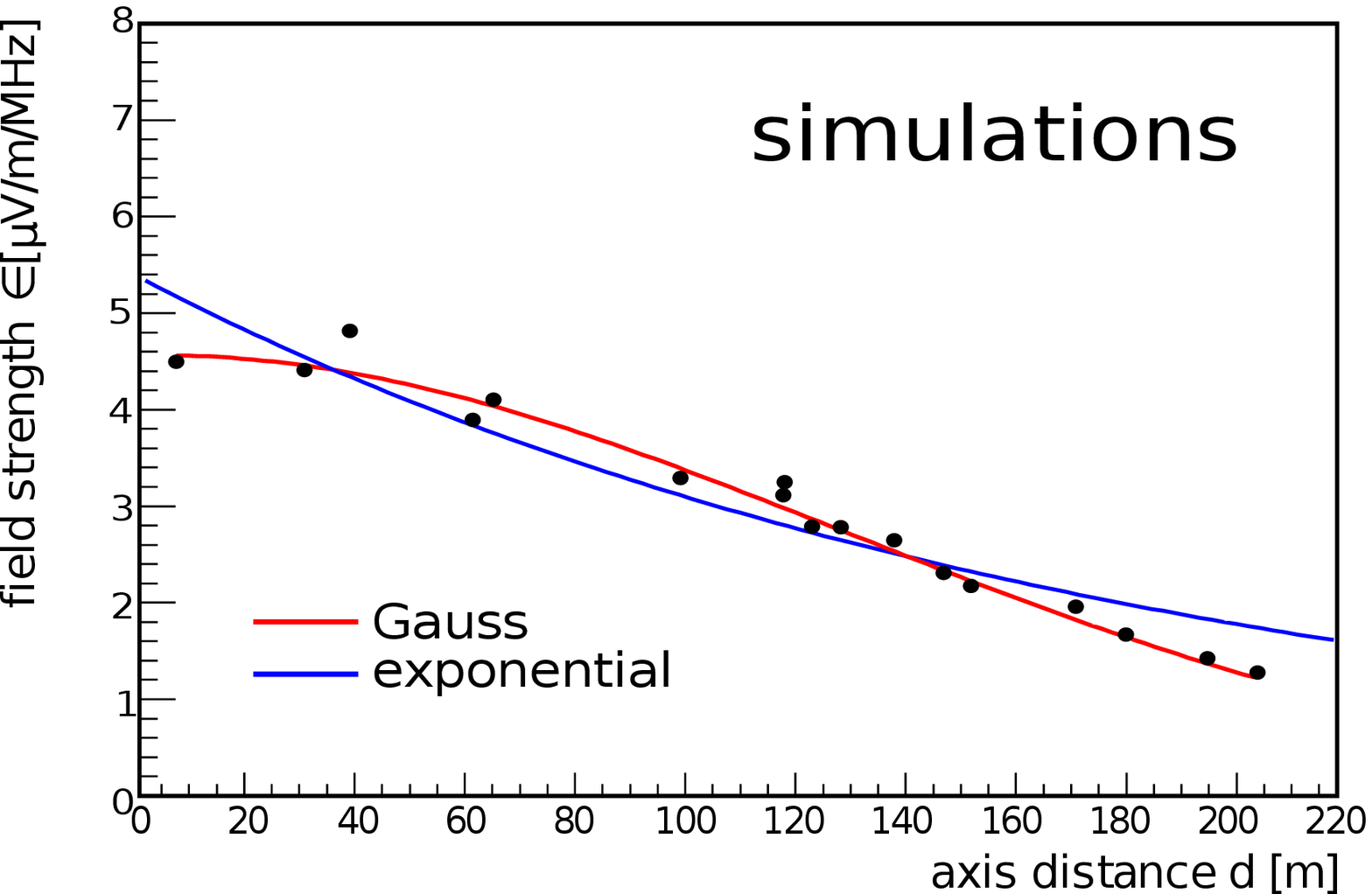}
\includegraphics[width=0.5\textwidth, height=0.35\textwidth]{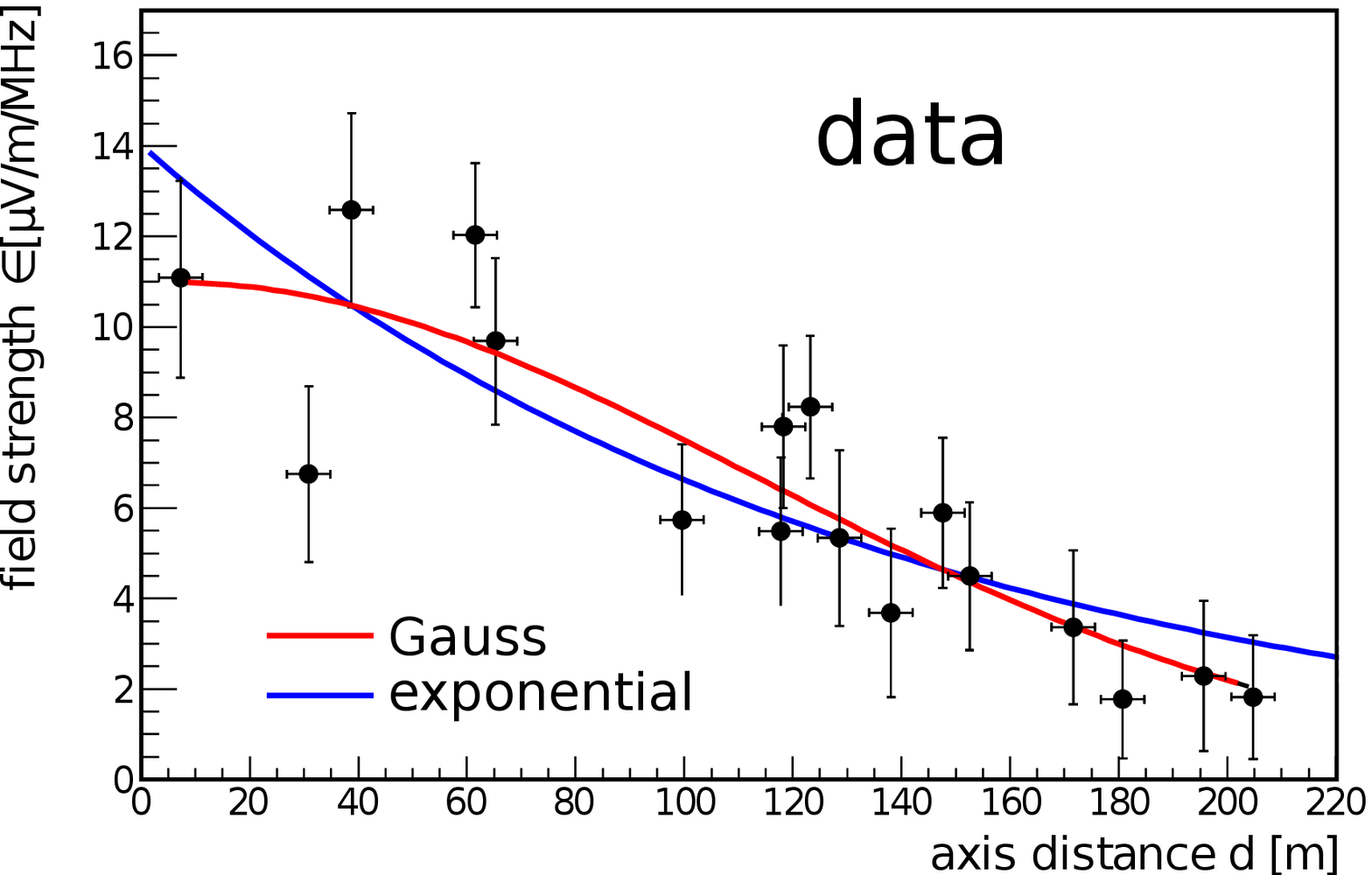}
\caption{Lateral distribution functions for an event in the selection used for the analysis. A Gaussian and an exponential function are used to fit the points. Top: CoREAS simulations. Bottom: LOPES data. The absolute amplitudes exhibit a known scale mismatch between LOPES data and CoREAS simulations (see text and ref.\ \cite{Schroeder_2013}).} \label{fig_lpLDF_gfit}
\end{figure}

\section{The lateral distribution function}
\label{sec_ldf}

The lateral distribution function (LDF) describes the measured (or simulated) electric field as a function of the observer position with respect to the shower axis. Due to the superposition of geomagnetic and charge-excess radiation in radio emission from extensive air showers, the LDF can exhibit significant asymmetries \cite{Huege_icrc13_ht}. Ideally, therefore a two-dimensional LDF taking into account the observer azimuth angle in addition to the axis distance should be fitted to the measured radio data. In the case of LOPES, however, the limited number of antennas sampling the true LDF led us to apply a simpler approach. Instead of an asymmetric two-dimensional LDF, we apply a one-dimensional LDF to our data, thereby effectively averaging out the asymmetries in the true LDF. Depending on the coverage in the sampled antenna locations, a per-event amplitude shift can be introduced by this procedure. This effect possibly leads to increased scatter in the fitted amplitudes, which is, however, implicitly accounted for in the following analysis. Another effect of the one-dimensional averaging could be an overall bias in the calibration of radio-amplitudes to primary particle energies determined by KASCADE. To quantify these two effects, we compared simulations including the charge-excess asymmetry with special simulations in which the charge-excess contribution has been removed. The result of this dedicated study are: The use of a one-dimensional LDF increases the scatter observed in the amplitude-energy correlations by an amount that is negligible with respect to other uncertainties, in particular LOPES experimental uncertainties. For the data set as a whole a systematic bias of 4.4\% in the estimated average amplitudes arises, which we account for in the following energy correlation analysis by adopting an extra systematic uncertainty of 5\%, well-below our global experimental calibration scale uncertainty of 35\%.

The uncertainty of the measured electric field amplitudes we fit with our one-dimensional LDF is taken as the root sum of squared uncertainties due to the individual antenna calibration (5~\%) and due to the noise \cite{Schroeder_thesis}. The distance of the antennas from the shower axis is calculated in the shower-plane coordinate system with a correlated uncertainty which depends on both the KASCADE precision ($\sim$~4~m \cite{Antoni_2003}) and the geometry reconstruction accuracy ($<$~0.3$^{\circ}$). In contrast, the CoREAS electric field at each observer position is assumed to be exact, thus no uncertainty is associated with the pulse amplitude.

The CoREAS simulations
are able to reproduce the complexity of the measured radio LDF, including asymmetries and flattening due to Cherenkov-like time compression effects. Even at the distances probed by the LOPES experiment, this complexity is clearly visible \cite{Schroeder_2013,lopesLDF}. Since only events with a core position relatively close to the antenna array are selected (sec.\ \ref{sec_evSelection} and fig.\ \ref{fig_map_selection}), the flattening of the radio lateral distribution close to the reconstructed core becomes an important aspect for our investigation. 
In previous LOPES analyses, we have used an exponential function to fit data and simulations of LOPES events \cite{Schroeder_2013,lopesLDF}. Here, we prefer to use a one-dimensional Gaussian function
  \begin{equation}
  \hspace{2cm}
\epsilon (d) \simeq \epsilon_{\mathrm{G}} \exp \left(\frac{(d - b)^{2}}{2c^{2}} \right)
   \label{expf}
   \label{eqLDF}
\end{equation}
with $\epsilon_{\mathrm{G}}$ [$\upmu$V/m/MHz], $b$ [m], and $c$ [m] as 
free parameters to fit the LOPES LDF. While the average LOPES event can be fitted by either a one-dimensional exponential or a Gaussian with similar quality, there are two advantages in using the Gaussian. For one, there are some events for which the Gaussian can fit the data well, while the exponential gives a considerably worse fit. Second, and more importantly, we develop our analysis method on the basis of CoREAS simulations, and for simulated events the Gaussian fits considerably better than an exponential. We illustrate this with a comparison of a Gaussian and an exponential fit to data and simulations in fig.\ \ref{fig_lpLDF_gfit}. We stress that the analysis presented here can be (and has also been) performed on the basis of an exponential LDF, and the results agree within the systematic uncertainties quoted here. However, for the above reasons, we prefer to use a Gaussian LDF.

Besides the depth of the shower maximum (X$_{\mathrm{max}}$), 
the air-shower inclination influences the slope of the radio LDF. This is related as well to a geometrical effect: near-vertical events constitute a radio source closer to the observer at ground level than inclined showers, and therefore are characterized by a steeper slope of the radio LDF. Five zenith angle bins are considered separately
to reduce the dependence on the inclination and to focus on the effects connected to the depth of shower maximum of the showers. These zenith angle bins are chosen such that they cover the same solid angle.

\begin{figure}[ht!]
\centering
\includegraphics[width=0.5\textwidth]{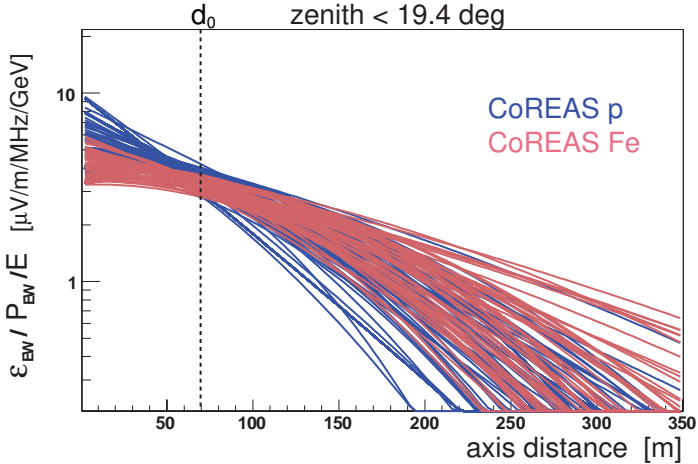}
\caption{Normalized CoREAS lateral distribution function fits using a one-dimensional Gaussian for the events with zenith angles smaller than 19.4$^{\circ}$, simulated once as proton (blue), once as iron (light-red) primaries. The dotted line represents the distance d$_{0}$ at which the RMS-spread of the fitted normalized amplitudes is minimal.} 
\label{fig_ldfE}
\end{figure}
\begin{figure}[ht]
\includegraphics[width=0.5\textwidth]{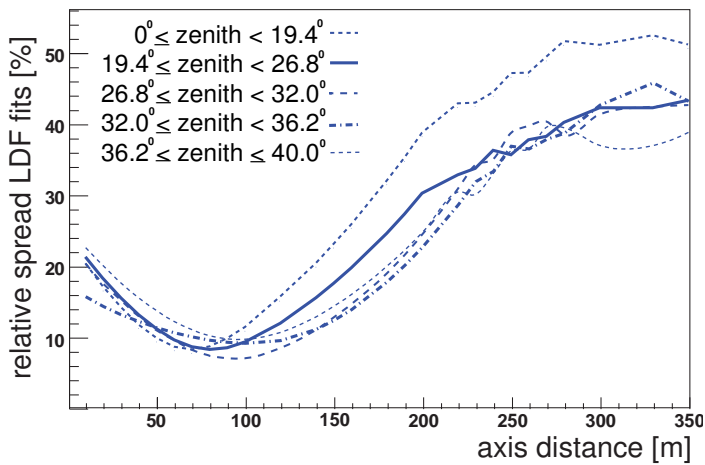}
\caption{Relative RMS-spread profile for the CoREAS LDF fits as a function of distance from the shower axis, for the different zenith angle bins. } 
\label{fig_rmsLDFall}
\end{figure}

\section{The slope method}

We first investigate the slope method using the CoREAS simulations of the LOPES events. The procedure aims to compare the LDF of several events with different primary energies and arrival directions, thus normalizations for the radio amplitudes are required. With the well-founded assumption that the leading contribution to the radio emission from air showers is of geomagnetic origin ($>$~85~\%) \cite{Huege_icrc13,HornefferICRC2007}, the East-West channel amplitudes for each given event are corrected for $\mathrm{\overrightarrow{P}_{ew}}$, the east-west component of the Lorentz vector ($\mathrm{\overrightarrow{v}}\times\mathrm{\overrightarrow{B}}$ ), with $\mathrm{\overrightarrow{v}}$ the arrival direction and $\mathrm{\overrightarrow{B}}$ the geomagnetic field vector. (Due to detection threshold effects, there are no events in the LOPES data set with a shower axis parallel to the magnetic field; the smallest geomagnetic angle is 15$^\circ$, corresponding to a $\sin(\alpha)$ of 0.26, whereas 
the mean $\sin(\alpha)$ is 0.7.) In addition, we normalize the electric field amplitudes by the Monte Carlo true energy of each event, which has been adopted as the energy reconstructed by KASCADE(-Grande) for the underlying LOPES event.

The normalized lateral distribution function fits for the events in the first zenith angle bin (zenith$<$19.4$^{\circ}$) are presented here as an example (fig.\ \ref{fig_ldfE}). This figure already illustrates that the spread of the obtained radio LDF fits drastically varies as a function of the distance from the shower axis. The  spread is quantified in fig.\ \ref{fig_rmsLDFall}, where the profile of the relative RMS-spread for the LDF fits is shown for each of the zenith angle bins. The spread reaches from $\approx 10$~\% at distances of 70-100~m from the shower axis up to more than 50~\% at distances larger than 300~m. Simulations \cite{REAS2,Palmieri_arena12} previously predicted the existence of a peculiar distance (d$_{0}$) where the radio amplitudes do not carry information on X$_{\mathrm{max}}$ and the influence of shower-to-shower fluctuations is minimal. Also here, we observe the minimum spread of the LDF fits at a peculiar distance d$_{0}$.
The value of d$_{0}$ as well as the value of the minimum RMS-spread vary with the zenith angle of the air shower and are summarized in table \ref{tab_$flat$ region}.
For LOPES events, d$_{0}$ lies in the range 70-100~m from the shower axis.

\begin{table*}[t]
\centering
\begin{tabular}{c|c|ccc}
  & & CoREAS sim.   \\ 
\hline  $\Delta \theta$& entries   &  d$_{0}$ [m] & RMS-spread [$\%$] & total uncertainty [$\%$]\\ 
\hline  
   $~\,0^{\circ}~-19.4^{\circ}$    &   53 & 70   & $~8.7$ & $13.3$ \\ 
   $19.4^{\circ}-26.8^{\circ}$  &   48 & 80   & $~8.1$ & $12.9$ \\ 
   $26.8^{\circ}-32^{\circ}~\,$    &   45 & 90   & $~7.6$ & $12.5$ \\ 
   $32^{\circ}~\,~-36.2^{\circ}$    &   36 & 100  & $~9.4$ & $13.7$ \\ 
   $36.2^{\circ}-40^{\circ}~\,$    &   23 & 100  & $10.0$ & $14.1$ \\
\hline 
\end{tabular} 
\caption{Distances d$_{0}$ for the five zenith angle bins as determined from CoREAS simulations. For each bin, the RMS spread for the LDF fits at d$_{0}$ of the different events is given, as well as a conservative estimate of the total uncertainty (computed as quadratic sum of the given RMS spread and the spread of individual points around the fits between $70\,$m and $100\,$m).}
\label{tab_$flat$ region}
\end{table*}  

\begin{table*}[t]

\begin{tabular}{c|c|c|cc|cc}

   &  &  & CoREAS sim.&  & LOPES data &  \\
 
\hline &  & predicted & & slope par. $k$&&slope par. $k$ \\ 
  $\Delta \theta$ & events & d$_{0,\mathrm{CoREAS}}$ & $\sqrt{\mathrm{RMS\small{_{spread}}}^{2}+\mathrm{20\%}^{2}}$  & $\pm$ stat. uncert.&  RMS\small{$_{\mathrm{spread}}$}& $\pm$ stat. uncert.\\ 
 &  & [m] & [\%]&\small{[GeV~m~MHz/$\upmu V$]} & [\%] &\small{[GeV~m~MHz/$\upmu V$]}\\

\hline  
    $~\,0^{\circ}~-19.4^{\circ}$   &   53 & 70  &$21.8$ & $0.299 \pm 0.003$&$18.8 $& $ 0.134 \pm 0.005$ \\ 
    $19.4^{\circ}-26.8^{\circ}$ &   48 & 80  &$21.6$ & $0.328 \pm 0.003$&$23.1 $& $ 0.151 \pm 0.006$ \\ 
    $26.8^{\circ}-32^{\circ}~\,$   &   45 & 90  &$21.4$ & $0.358 \pm 0.004$&$24.6 $& $ 0.141 \pm 0.006$ \\ 
    $32^{\circ}~\,-36.2^{\circ}$   &   36 & 100 &$22.1$ & $0.382 \pm 0.005$&$20.3 $& $0.142 \pm 0.007$ \\ 
    $36.2^{\circ}-40^{\circ}~\,$   &   23 & 100 &$22.4$ & $0.398 \pm 0.006$&$25.6$ & $0.137 \pm 0.008 $\\
\hline 
\end{tabular} 
\caption{Comparison between the CoREAS prediction and the LOPES measurements for the RMS-spread at the distance d$_{0}$, and for the slope paramters $k_{\mathrm{CoREAS}}$ and $k_{\mathrm{LOPES}}$ determined with linear fits (see figs. \ref{fig_energy} and \ref{fig_energyLOPES} for the first zenith angle bin). In addition to the reported statistical uncertainties of the $k$-parameters, an additional 5\% systematic uncertainty has to be taken into account because of the adoption of a symmetrical one-dimensional LDF. Since the CoREAS results are based on the KASCADE energy as Monte Carlo true input without uncertainty, but the LOPES results implicitly contain the KASCADE energy uncertainty (approx.\ $20\,\%$), we quadratically added $20\,\%$ to the CoREAS RMS spread.}
\label{tab_slopePar_linearFit}
\end{table*} 
 

\section{Primary energy reconstruction } 
\label{sec_energy}
\begin{figure}[t]
\centering
\includegraphics[width = 0.5\textwidth, height = 0.38\textwidth, angle=0]{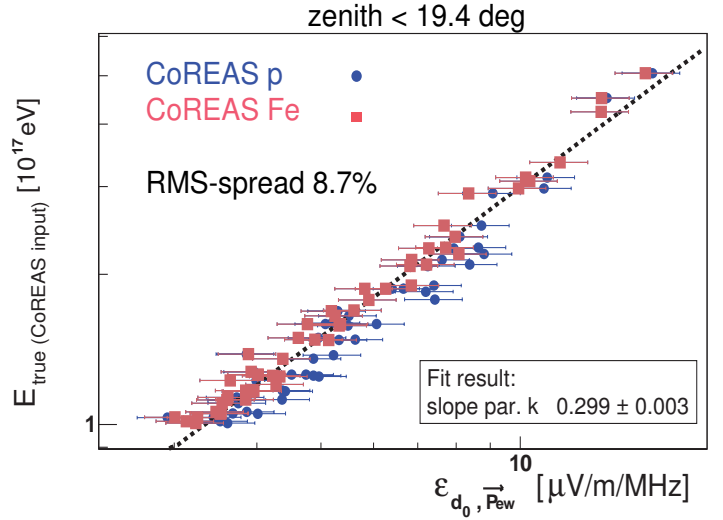}
\caption{Linear correlation between the Monte Carlo true primary energy and the CoREAS radio pulse amplitude at the distance d$_{0}$, with $k$ the free parameter of the fit. The radio amplitude is normalized by the east-west component of the Lorentz force vector. The RMS-spread is $\sim$~8.7~\%.} \label{fig_energy}
\end{figure}
\begin{figure}[t]
\centering
\includegraphics[width = 0.5\textwidth, height = 0.38\textwidth, angle=0]{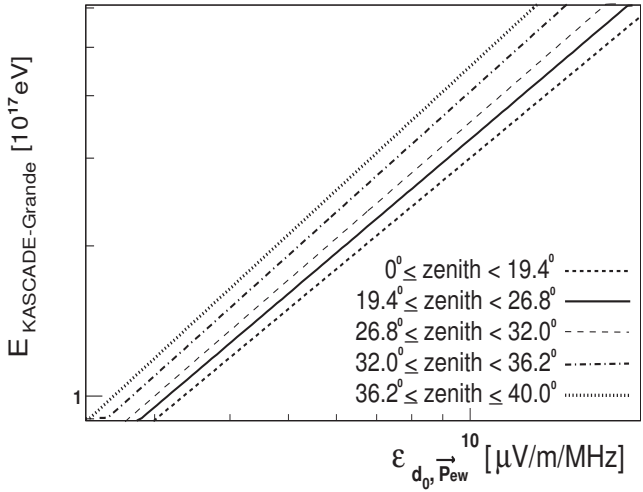}
\caption{Linear fits for the energy -- radio amplitude in d$_{0}$ correlation (double-log scale), for all the zenith angle bins. The slope of each fit clearly depends on the inclination of the air shower.} \label{fig_fitsenergy}
\end{figure}
The radio amplitude at the peculiar distance d$_{0}$, barely affected by shower-to-shower fluctuations, provides direct information on the shower energy \cite{REAS2,Palmieri_arena12}.

\subsection{CoREAS simulations}
To illustrate this first on the basis of simulations, we take the east-west component of the electric field vector predicted by CoREAS simulations of the LOPES events, apply the digital bandpass-filter, fit the Gaussian LDF, read off the radio pulse amplitude at the distance d$_{0}$ and normalize this amplitude with the east-west component of the Lorentz force vector adequate for the arrival direction (i.e. $\mathrm{\overrightarrow{P}_{ew}}$). We then plot this normalized amplitude against the Monte Carlo true energy of the air shower, as shown in fig.\ \ref{fig_energy} for the events in the first zenith angle bin. 

A simple linear correlation with the total primary energy becomes apparent: eq.\ \ref{enFlat_ch7} with $k$ 
[GeV~$\cdot$~m~$\cdot$~MHz~/$\upmu$V] as free parameter is used to fit the filtered radio amplitudes at the distance d$_{0}$ for proton and iron simulated events.
\begin{equation}
\label{enFlat_ch7}
\mathrm{energy} = k\cdot(\epsilon_{\mathrm{d_{0}}}/\vert\mathrm{\overrightarrow{P}_{ew}}\vert) 
\end{equation} 

A measurement of the radio pulse amplitude at the peculiar distance d$_{0}$ thus allows us to directly reconstruct the energy of the cosmic ray primary. The intrinsic uncertainty of this method is composed of two contributions: the RMS-spread of the amplitude at d$_{0}$ around the linear correlation with the energy (approx.\ $9\,\%$), and the scatter of the individually measured radio amplitudes at d$_{0}$ (approx.\ $10\,\%$). Of course, both quantities are correlated, but as a conservative estimate we added both quadratically and estimate the total energy uncertainty to approximately $13\,\%$ (see table \ref{tab_$flat$ region}). This might be improved by explicitly taking the asymmetries introduced by the Askaryan effect into account in the fitting procedure, since they are the main source of the scatter. However, such an improvement would be below our testing power which is limited by the energy precision of KASCADE-Grande ($\sim 20\,\%$).

On closer look, a systematic shift between the normalized radio amplitudes at the distance d$_{0}$ is visible between the two different primaries (fig.\ \ref{fig_energy}). The radio-amplitudes for proton primaries are slightly higher than those for iron primaries. This is related to the fact that the radio emission is connected only to the electromagnetic component of the air shower. Depending on the primary type, a slightly different fraction of the total primary energy is transferred to the electromagnetic and non-electromagnetic components of the air shower, thus changing the amplitudes of the radio signal \cite{REAS2,HeitlerMatthews}. 

We stress here that this is not a ``problem'' of the radio detection technique. It simply signifies that radio emission purely probes the energy in the electromagnetic component of an air shower, with a high intrinsic precision. Thus, combining a radio detector with a technique sensitive to the muonic component of air showers would be a particularly compelling approach potentially giving very detailed information on the cosmic ray composition. This feature also makes the interpretation of radio measurements less sensitive to hadronic interaction model uncertainties, as the electromagnetic component of air showers is generally well-understood.

The linear correlations for all the zenith angle bins are shown in fig.\ \ref{fig_fitsenergy}. The values for the slope parameter $k$ are reported in table \ref{tab_slopePar_linearFit}. Table \ref{tab_slopePar_linearFit} also denotes the RMS-spread around the linear fit --- an estimator of the achievable energy precision --- expected when taking into account the KASCADE energy uncertainty: The simulations use the KASCADE energy as Monte Carlo true input and thus are not affected by any uncertainty of this energy. In contrast, for LOPES measurements, the KASCADE energy uncertainty has a significant effect on the results (thus, there are y-error bars in fig.~\ref{fig_energyLOPES} (top), but not in fig.~\ref{fig_energy}). Consequently, for a comparison between simulations and data the KASCADE energy uncertainty has to be added to the values derived from simulations. For the data set and energy range used here, this uncertainty is not known with very high precision, and is estimated to be 
approximately $20\,\%$. We thus quadratically add a constant $20\,\%$ to the RMS spread of the simulations. As described in section \ref{sec_ldf}, the $k$-parameters reported here have an additional 5\% systematic uncertainty due to the use of a symmetric one-dimensional LDF function to estimate the per-event amplitude at the characteristic distance.

A deeper look at the absolute values of the $k$-parameters reveals that they increase with increasing zenith-angle. This is due to the evaluation of the fit at the specific characteristic distance $\mathrm{d}_{0}$, which increases with zenith angle. As $\mathrm{d}_{0}$ increases, the average radio amplitude decreases, and thus the $k$-parameter increases. If the correlations are evaluated at a fixed lateral distance for all zenith angle bins, the $k$-parameters become consistent within uncertainties.

\subsection{LOPES measurements}
\label{sec_energy_lopes}
\begin{figure}[t!]
\centering
\includegraphics[width = 0.50\textwidth,angle=0]{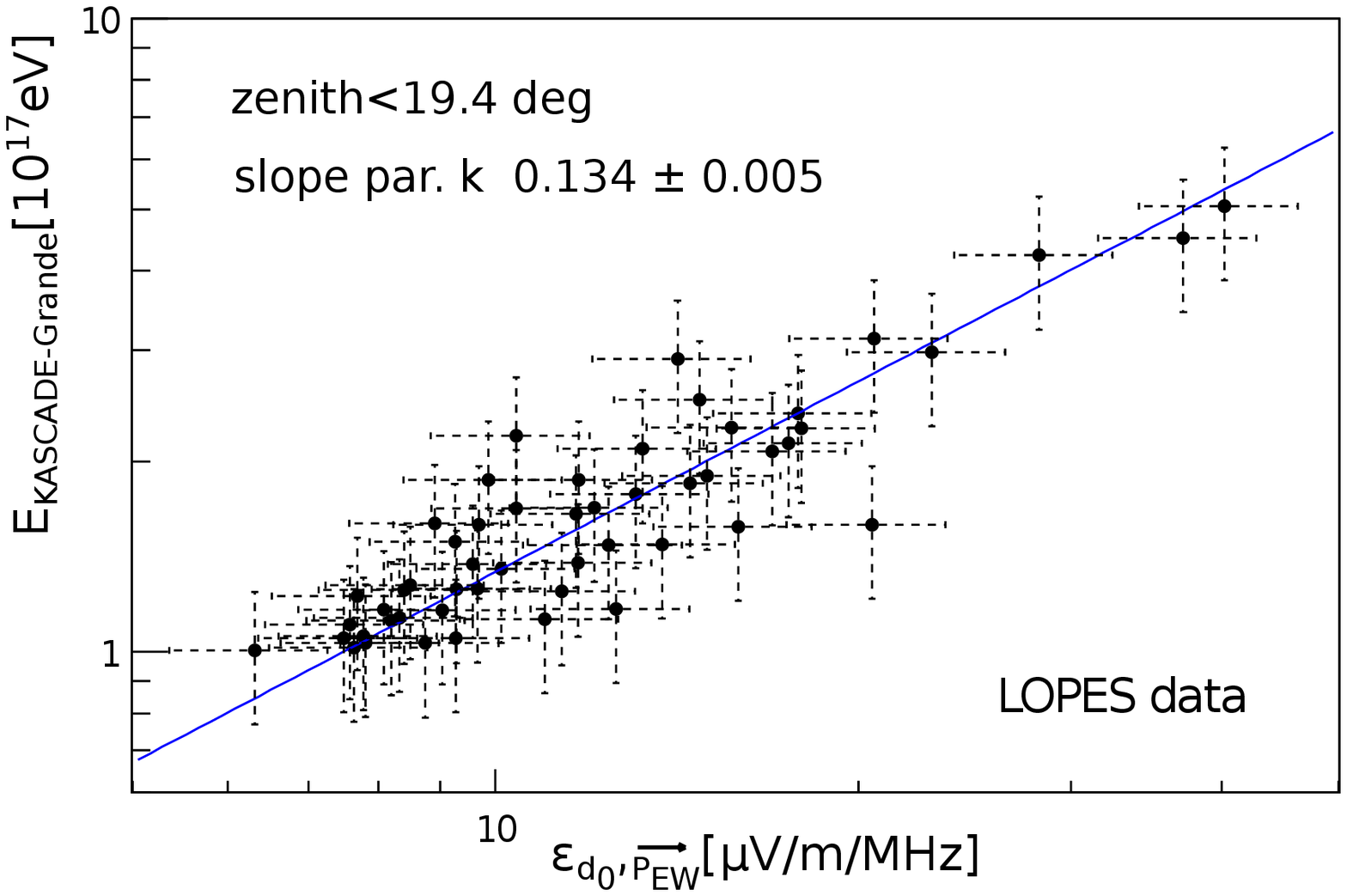}
\includegraphics[width = 0.50\textwidth]{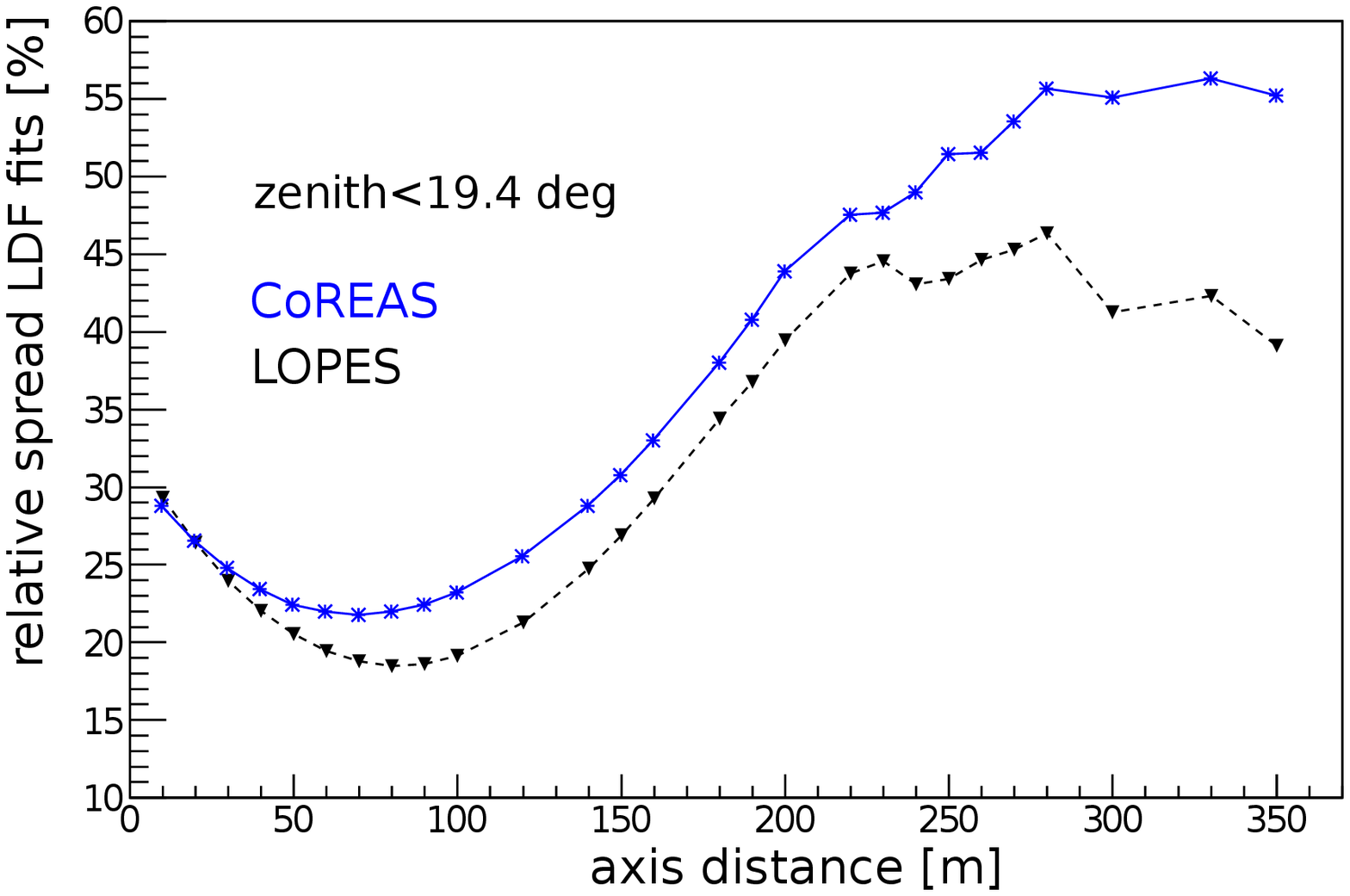}
\caption{Top: Linear correlation of the KASCADE(-Grande) reconstructed primary energy and the normalized LOPES measured radio pulse in d$_{\mathrm{0}}$, for zenith angles less than 19.4$^{\circ}$. The RMS-spread is bf 18.8~\%. Bottom: RMS-spread from the energy-fit for zenith angle less than 19.4$^{\circ}$, computed at several distances from the shower axis. With the LOPES measurements (black triangles) the minimum is found at 80~m. The blue stars represent the CoREAS predictions (fig.\ \ref{fig_rmsLDFall} for the same events, adding $20\,\%$ for the KASCADE energy uncertainty). }
 \label{fig_energyLOPES}
\end{figure}


We now apply the energy reconstruction to LOPES data. The LOPES-measured radio lateral distributions for each event are fit with the Gaussian function. The fit value at the distance d$_{\mathrm{0}}$ is used as the energy estimator. We adopt the values for the distances d$_{0}$ at which we evaluate the radio amplitude in each zenith angle bin as those derived previously from the CoREAS simulations (table \ref{tab_$flat$ region}).

This yields a simple linear correlation between the measured radio pulse (at distance d$_{\mathrm{0}}$) and the energy reconstructed by KASCADE-Grande, as shown in fig.\ \ref{fig_energyLOPES} -top panel (zenith angles smaller than 19.4$^{\circ}$ taken as example).
The slope parameter $k$ for the linear fit in each zenith bin (reported in table \ref{tab_slopePar_linearFit}, right side) is not compatible within the uncertainties with the CoREAS predicted slope. This illustrates the still existing amplitude scale mismatch between CoREAS simulations and LOPES measured radio pulses \cite{Schroeder_2013}, which on average amounts to a factor of $\approx 2.5$. Further comparison between the $k$-parameters determined for simulations and LOPES data reveals that the deviation increases with zenith angle. We have performed an independent analysis on a different selection of LOPES events and compared the measured amplitudes at a distance of 100~m with the amplitudes predicted by CoREAS simulations. This cross-check confirmed that the amplitude discrepancy between LOPES measurements and CoREAS simulations increases with zenith angle. The reason for this discrepancy is currently unknown. It could be caused by a discrepancy between the simulated LOPES antenna gain pattern and the actual antenna gain pattern, in spite of efforts to cross-check our antenna pattern simulations with dedicated calibration measurements \cite{Nehls2008}. Another explanation could be a deterioration of the KASCADE-based (not KASCADE-Grande-based) energy reconstruction at energies well beyond $10^{17}$~eV \cite{Schroeder_2013}, where punch-through of the electromagnetic air shower component into the shielded muon detectors may start to matter, an effect which is obviously zenith-angle dependent. Finally, there could also be a problem in the CoREAS simulations, which should thus be compared with data from other experiments.

As a next step, we verify the existence of the peculiar distance d$_{\mathrm{0}}$ in the LOPES measurements. At the distance d$_{\mathrm{0}}$ the correlation between the energy and the radio amplitude exhibits the lowest uncertainty. Thus, we use again the RMS-spread around the linear fit (fig.\ \ref{fig_energyLOPES}-top panel) of the measured data, but now evaluate the RMS-spread over the full range of relevant lateral distances d$_{\mathrm{i}}$ from the shower axis. The RMS-spread around each linear fit is again referred to as the precision for the primary energy reconstruction achievable at each distance d$_{\mathrm{i}}$. The RMS-spread obtained at these several distances, for zenith angles smaller than 19.4$^{\circ}$, is shown in fig.\ \ref{fig_energyLOPES}, bottom panel (black triangles).

The measurements show a minimum at 80~m from the shower axis, with an RMS value of 19~\% (and slightly higher values for the other zenith angle bins, cf.~table \ref{tab_slopePar_linearFit}). Although there is a slight deviation of the measured d$_{\mathrm{0}}$ value of 80~m with respect to the CoREAS-predicted value of 70~m, one may notice that the RMS-spread for the measurements differs by only 1 percentage point between 60~m and 100~m, as the minimum is rather broad. The shift between the predicted and the measured d$_{\mathrm{0}}$ is, therefore, not practically relevant, and the radio amplitude measured at the simulation-predicted distance d$_{\mathrm{0}}$ can still be used to determine the upper limit on the primary energy reconstruction precision with LOPES radio data.

The RMS-spreads of each of the linear fits of the LOPES measurements at the simulation-predicted distance d$_{0}$ are summarized in table \ref{tab_slopePar_linearFit}. These values contain the combination of the LOPES- and KASCADE-Grande energy uncertainties. For comparison with the simulations, an additional uncertainty of 20~\%, representative for the KASCADE-Grande energy uncertainty  \cite{Antoni_2003}, is added to the spread determined from the CoREAS simulations. The resulting spread is shown as well in fig.\ \ref{fig_energyLOPES}-bottom panel (blue stars) and in table \ref{tab_slopePar_linearFit}. In case of the first zenith angle bin, the spread determined from the simulations with an added 20 \% even seems to over-estimate the spread in the LOPES data. At least in this zenith angle range, the KASCADE energy resolution might therefore be actually better than 20 \%. The upper-limit on the energy reconstruction achieved with the LOPES measurements, in all the zenith angle bins, is only slightly higher than the uncertainty of the KASCADE-Grande energy resolution. 

The LOPES energy resolution is, therefore, very promising, and the application of similar procedures on experiments with a better reference energy resolution and radio data affected by a lower ambient noise background is highly suggested in order to properly verify the simulation-based expectations of a high intrinsic energy resolution of radio measurements.

\section{X$_{\mathrm{max}}$ reconstruction}
 \label{sec_Xmax}
\begin{figure}[t!]
\centering
\includegraphics[width = 0.46\textwidth]{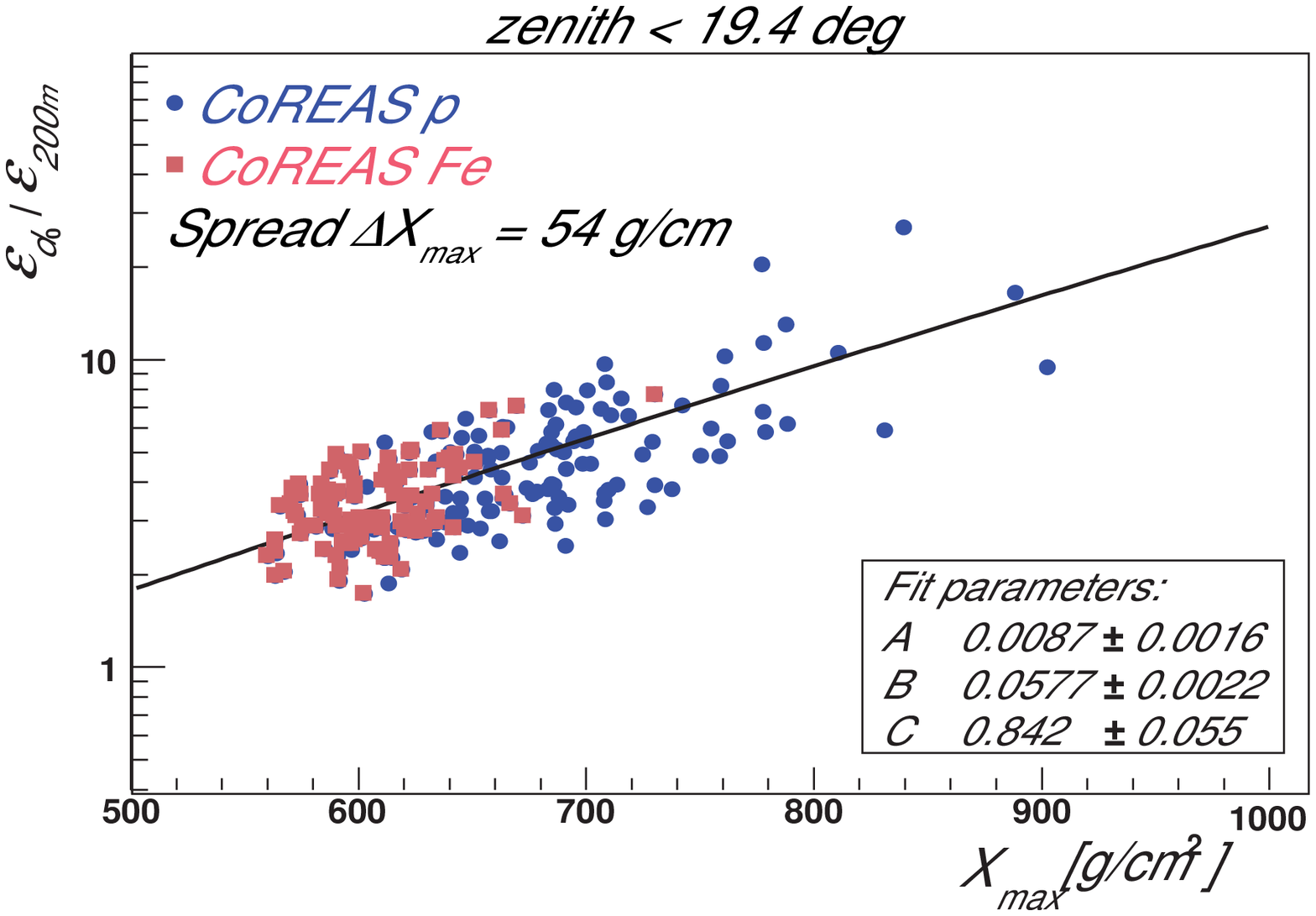}
\caption{Correlation between the true X$_{\mathrm{max}}$ from Monte Carlo simulations and the LDF slope of the CoREAS simulated radio amplitudes. The RMS-spread around the fit is $\sim$54~g/cm$^{2}$.}
\label{fig_Xmaxfit}
\includegraphics[width = 0.5\textwidth, height = 0.38\textwidth, angle=0]{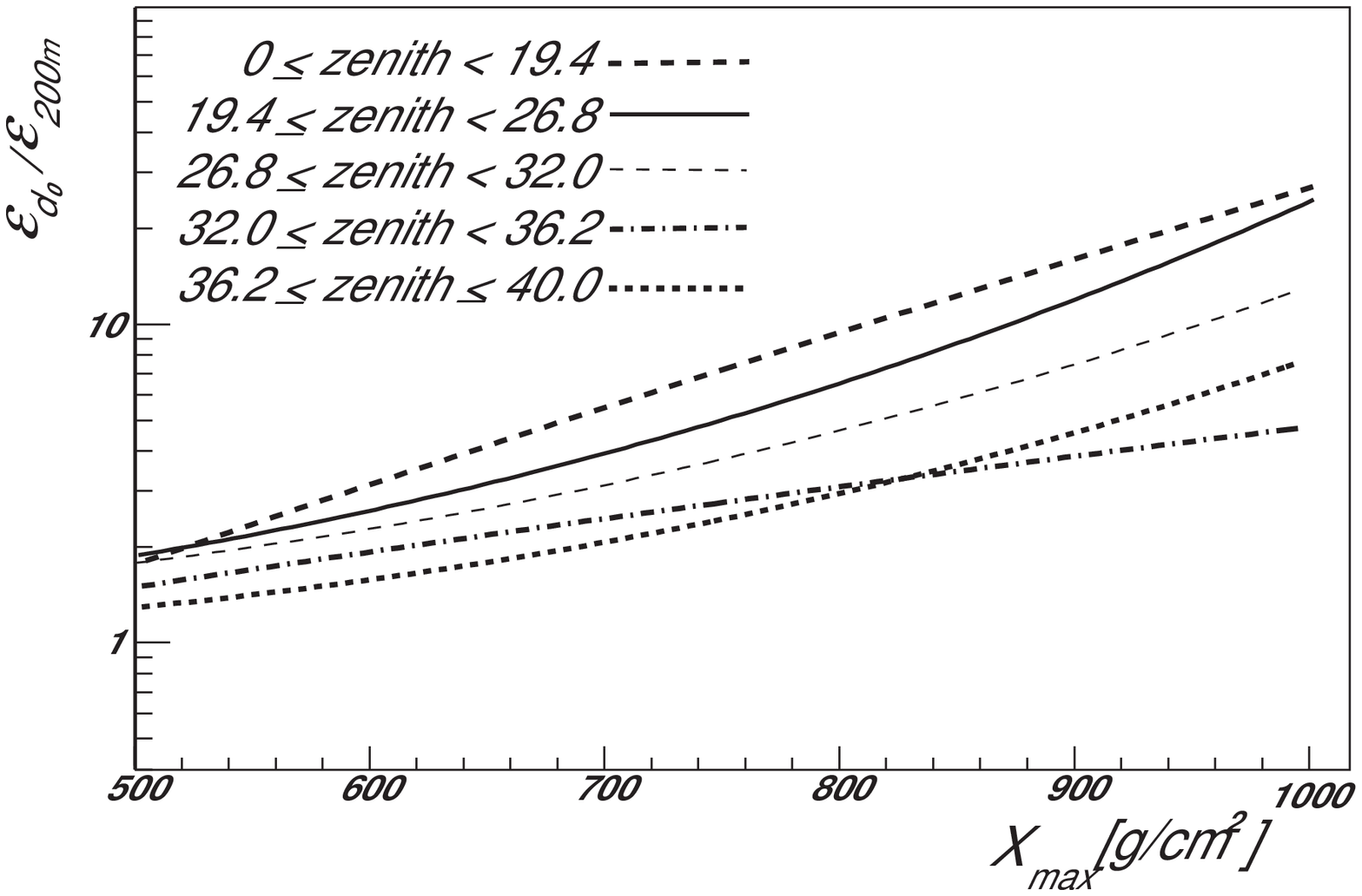}
\caption{Fits as in fig.\ \ref{fig_Xmaxfit} but for each zenith angle bin. The expected dependence of the X$_{\mathrm{max}}$--radio LDF slope correlation on the zenith angle is clearly visible.} 
\label{fig_Xmaxfits}
\end{figure}

\begin{table*}[tb!]
\begin{center}
\begin{minipage}[b]{1\textwidth}
\centering
\begin{tabular}{c|c|cccc}
             & &    &QGSJET II.03 &\& CoREAS    &\\
\hline               $\Delta \theta$ &entries &$A$ [cm$^{2}$/g] &  $B$        & $C$  & $\Delta$X$_{\mathrm{max}}$  \\ 
\hline  0.$^{\circ}$-19.4$^{\circ}$  & 53 *2    &0.0087~$\pm$~0.0016 & 0.0577~$\pm$~0.0022 & 0.841~$\pm$~0.055 & 54.0    \\ 
        19.4$^{\circ}$-26.8$^{\circ}$& 48 *2    &0.0060~$\pm$~0.0034 & 0.0835~$\pm$~0.0084 & 0.945~$\pm$~0.016 & 47.9 \\ 
        26.8$^{\circ}$-32$^{\circ}$  & 45 *2    &0.0014~$\pm$~0.0010 & 1.100~$\pm$~0.092 & 2.542~$\pm$~0.011 & 53.2   \\ 
        32$^{\circ}$-36.2$^{\circ}$  & 36 *2    &0.0053~$\pm$~0.0018 & 0.225~$\pm$~0.018 & 0.691~$\pm$~0.027 & 65.7\\ 
        36.2$^{\circ}$-40$^{\circ}$  & 23 *2    &0.0013~$\pm$~0.0002 & 0.98~$\pm$~0.52 &  2.8~$\pm$~2.0 & 47.9\\
\hline 
\end{tabular} 
\caption{X$_{\mathrm{max}}$-reconstruction parameters derived with the CoREAS simulations and their statistical uncertainties. In the rightmost column, the uncertainty on the X$_{\mathrm{max}}$ reconstruction (CoREAS) for each zenith angle bin is listed (see text). The values in this table have been derived by using and increased (double) statistics for the simulations.}
\label{xmax_sel2}

\end{minipage}
\end{center}
\end{table*}

The feature of the radio lateral distribution function which correlates best with the depth of shower maximum is the slope of the LDF fit. We first ``calibrate'' the relation of this slope with the depth of shower maximum on the basis of CoREAS simulations. Afterwards, we apply the found relation to LOPES measurements.

\subsection{CoREAS simulations}

We define the LDF slope $\epsilon_{\mathrm{ratio}}$ as the ratio of the radio amplitudes taken at the distance d$_{0}$ and at the fixed distance of 200~m ($\epsilon_{\mathrm{ratio}}$=$\epsilon_{\mathrm{d_{0}}}$/$\epsilon_{\mathrm{200m}}$). The latter value is chosen for the specific investigation on the LOPES events, since 200~m is a relatively large distance but is still in the range of the measurements. 

The X$_{\mathrm{max}}$ values for each event are taken directly from the Monte Carlo simulations, and are considered as the ``true'' values. 
 The ratio of the radio amplitudes and the depth of the shower maximum are plotted against each other in fig.\ \ref{fig_Xmaxfit}, for zenith angles less than 19.4$^{\circ}$. There is a clear correlation which we fit with eq.\ \ref{xm_eq}.
\begin{equation}
\hspace{1.5cm}
\epsilon_{\mathrm{ratio}} = B\ \mathrm{exp}\left[ \left(A \cdot X_{\mathrm{max}}\right)^{C}\right]
\label{xm_eq}
\end{equation}
This (more precisely the inverse) three-parameter function was already introduced in \cite{REAS2} and successfully applied to previous analyses \cite{Palmieri_arena12,Palmieri_thesis}. The function (eq.\ \ref{xm_eq}) appropriately describes the correlation for the events simulated with CoREAS, and it is adequate for all of the zenith bins. The $A$, $B$ and $C$ parameters, including the statistical uncertainties of the fit, are extracted separately for each zenith angle bin (table \ref{xmax_sel2}), and the corresponding fit results are shown in fig.\ \ref{fig_Xmaxfits}. This function has not been optimized for minimal correlation between fit parameters, which can lead to comparably good fits with more than one set of parameters, visible as fluctuations in the derived parameter values in table \ref{xmax_sel2}. This is fine as long as the resulting fits describe the distribution of data points correctly, which they do. 
However, in the future it would be worth investigating better fit functions with minimally correlated fit parameters, ideally parameters that can be associated directly with physical parameters of the underlying air shower.
To improve the determination of the three fitting parameters and to better constrain the fit over a wide range of X$_{\mathrm{max}}$ values, increased (doubled) statistics for the proton- and iron-simulated showers are used.

We performed the fit procedure for all zenith angle bins both without and with associating an uncertainty propagated from the LDF fits to the individual data points, and the results vary only insignificantly. Propagating the resulting fit uncertainties into an uncertainty for the reconstructed X$_{\mathrm{max}}$ values cannot be achieved by Gaussian error propagation due to the non-linear nature of the problem. One option to proceed would be to employ an involved procedure for the estimation of the uncertainty of the reconstructed X$_{\mathrm{max}}$ values, e.g.\ by bootstrapping methods. As we want to keep our analysis practical and are anyway limited by the lack of an independent X$_{\mathrm{max}}$ determination to compare the reconstructed values with, however, we choose instead to use the resulting spread in the distribution around the fit as an estimate for the uncertainty of the reconstructed X$_{\mathrm{max}}$ values. 

\begin{figure}[h!]
\centering
\includegraphics[width = 0.5\textwidth, height = 0.38\textwidth]{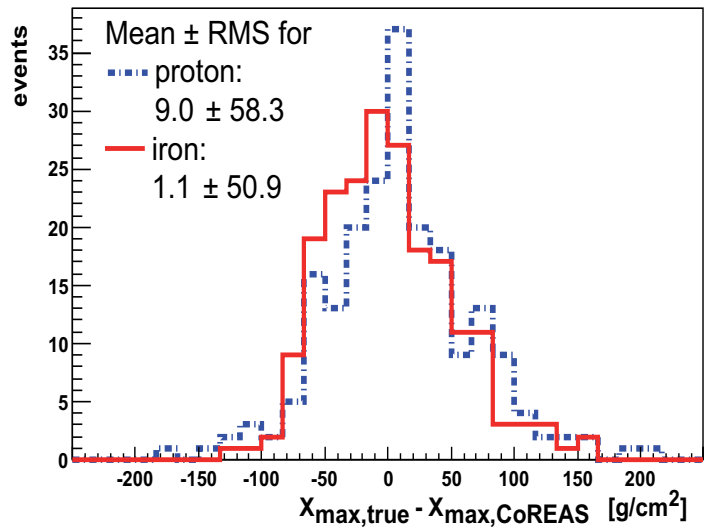}
\caption{Difference of the true Monte Carlo X$_{\mathrm{max}}$ and the value derived by applying the slope method on CoREAS simulations for proton showers (blue, dash-dotted line) and iron showers (red, solid line). The RMS spread of this difference is $58\,$g/cm$^{2}$ for proton showers and $51\,$g/cm$^{2}$ for iron showers. This spread constitutes an estimate for the intrinsic uncertainty of the slope method for the reconstruction of X$_{\mathrm{max}}$.} \label{fig_trueVsRecXmax}
\end{figure}

In other words, we calculate the typical X$_{\mathrm{max}}$ precision of the slope method as the spread of the difference between the true Monte Carlo X$_{\mathrm{max}}$ values and the X$_{\mathrm{max}}$ values reconstructed using the slope method on CoREAS simulations. Fig.\ \ref{fig_trueVsRecXmax} shows the corresponding distributions for proton showers (RMS-spread 58 g/cm$^{2}$) and iron showers (RMS-spread 51 g/cm$^{2}$) for the complete selection of events. In table \ref{xmax_sel2}, the combined spread for proton and iron showers is denoted as $\Delta$X$_{\mathrm{max}}$ for the individual zenith angle bins. For the CoREAS simulated events with doubled statistics, and considering the complete zenith angle range, 
the RMS-spread is in the range of 45-65~g/cm$^{2}$.

It is important to note that the results presented here apply to the specific situation of LOPES, in particular the altitude of the LOPES site, the specific observing frequency band used \cite{REAS2}, the energy range probed, and the dimension of the antenna array. These parameters all affect the amplitude ratio and consequently play a role in the determination of the X$_{\mathrm{max}}$ calibration curves. Indeed, for inclined showers, larger amplitude ratios may be obtained when the far-away distance used to calculate the ratio is taken well beyond the 200~m distance used for this analysis and imposed, here, by the dimension of the LOPES array.

\subsection{LOPES measurements}
\label{Xmaxreconstr}
\begin{figure}[th!]
\centering
\includegraphics[width = 0.5\textwidth, angle=0]{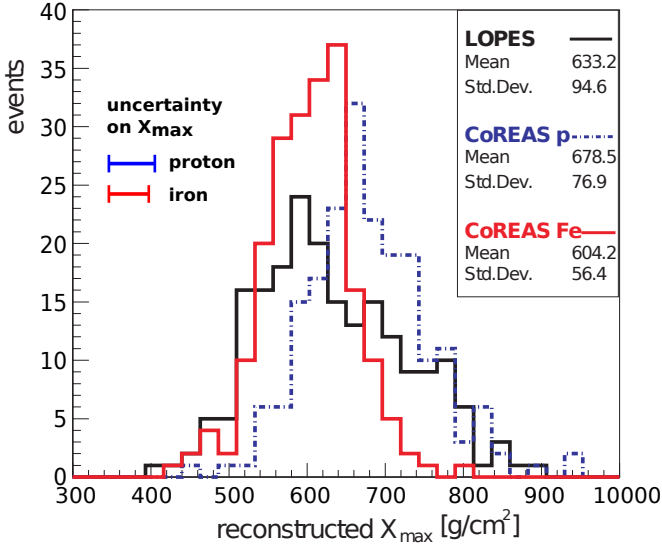}
\caption{X$_{\mathrm{max}}$ distribution reconstructed with the slope method for the LOPES measurements (black) and for the CoREAS proton (dashed-blue) and iron (light-red) simulations of the same events, for the complete zenith angle range of 0-40$^{\circ}$. The indicated uncertainty on X$_{\mathrm{max}}$ is derived from fig. \ref{fig_trueVsRecXmax}.} \label{fig_xmax}
\end{figure}
\begin{figure}[h!]
\centering
\includegraphics[width = 0.5\textwidth, height = 0.38\textwidth]{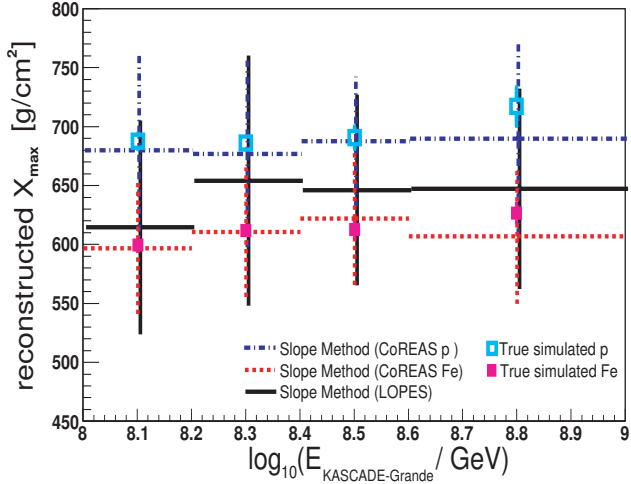}
\caption{The energy dependence of the reconstructed LOPES X$_{\mathrm{max}}$ distribution (black), compared with the CoREAS values for proton (blue-dashed line) and iron (red-dotted line) primaries, and with the Monte Carlo values for proton (light-blue marker) and iron (light-red marker) primaries. } \label{fig_xmaxlge}
\end{figure}

The three-parameters function (eq.\ \ref{xm_eq}) is now applied to reconstruct the X$_{\mathrm{max}}$ values from LOPES measurements, using the CoREAS-derived $A$, $B$ and $C$ parameters listed in table \ref{xmax_sel2} for the appropriate zenith angle bin of each individual event.

In fig.\ \ref{fig_xmax}, we show the resulting distribution of X$_{\mathrm{max, LOPES}}$ values in the black histogram, with 633.2~$\pm$~94.6~g/cm$^{2}$ as the mean and the standard deviation of the distribution. A comparison with the CoREAS proton (blue dotted line) and iron (red line) prediction is shown as well: X$_{\mathrm{max, CoREAS, P}}$~=~678.5~$\pm$~76.9~g/cm$^{2}$ and X$_{\mathrm{max, CoREAS, Fe}}$~=~604.2~$\pm$~56.4~g/cm$^{2}$. These values are derived by applying eq.\ \ref{xm_eq} to the (CoREAS) simulated events, or, in other words these are the values on the fit in fig.\ \ref{fig_Xmaxfit} of the corresponding simulated LDF slopes.

The standard deviation value for the LOPES distribution of $\sim$95~g/cm$^{2}$ can be considered itself as an upper-limit for the LOPES precision on X$_{\mathrm{max}}$, although it does not explicitly account for systematic uncertainties, e.g., scale uncertainties due to modelling the atmosphere or due to hadronic interaction models used for the shower simulations.

The reconstructed values for X$_{\mathrm{max,LOPES}}$ are compatible with the expectations. Compared to previous analyses \cite{Palmieri_arena12,Palmieri_thesis}, in which the simulations used did not include the refractive index of the atmosphere, a much improved result is now obtained. We can definitively affirm that the systematic shift observed previously in \cite{Palmieri_arena12} was caused by the discrepancy between the LDF slopes predicted by older simulation codes and the measured LDF slopes.

In addition, we examine the energy dependence of the X$_{\mathrm{max}}$ distribution (fig.\ \ref{fig_xmaxlge}), integrated over the whole zenith angle range: Each cross represents the mean value and the standard deviation of the X$_{\mathrm{max}}$ distributions, for the slope method applied to LOPES measurements and to CoREAS simulations. Additionally, the mean and the error of the mean (standard deviation/$\sqrt{\mathrm{events}}$) for the true Monte Carlo X$_{\mathrm{max}}$ are shown in each energy bin. Good consistency with the expectations is achieved in all energy bins.

\begin{figure*}[ht!]
\centering
\setlength\fboxsep{0pt}
\setlength\fboxrule{0.pt}
\fbox{
\includegraphics[width = 0.7\textwidth]{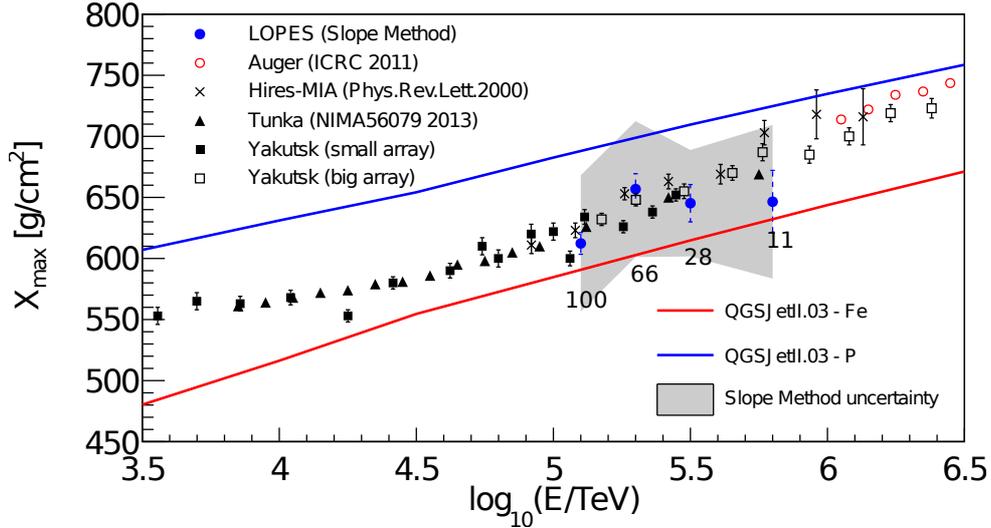}}
\caption{Mean X$_{\mathrm{max}}$ as a function of the primary cosmic ray energy. The blue points correspond to the LOPES-reconstructed values, using the slope method. The errors are calculated as (standard deviation/$\sqrt{\mathrm{events}}$), and the numbers indicate our statistics in each energy bin. The lines refer to the Monte Carlo expectations for pure iron (red) and pure proton (blue) composition, using the QGSJetII-03 hadronic interaction model. Results from other experiments are indicated by markers (please see legend). } \label{fig_xmspectra}
\end{figure*}

Next, we directly compare the mean LOPES X$_{\mathrm{max}}$ values reconstructed for the four energy bins with the results of other experiments (fig.\ \ref{fig_xmspectra}). In the figure, the solid lines represent the Monte Carlo expectations for QGSJetII-03 for pure iron-simulated (red) and pure proton-simulated showers. The uncertainty indicated in this figure on the LOPES average X$_{\mathrm{max}}$ values is the statistical uncertainty, i.e.\ standard deviation/$\sqrt{\mathrm{events}}$. The intrinsic uncertainty of the slope method, as determined by the spread between the reconstructed and Monte Carlo true values, is indicated as a shaded band. No additional analysis on the systematics is carried out. 

In summary, the results of the LOPES X$_{\mathrm{max}}$ determination are self-consistent and plausible. Nevertheless, we point out that the LOPES uncertainties are too large for any conclusion concerning the mass composition.

\section{Discussion}
In the following discussion, willful and imposed limitations which had influence on the slope method analysis of the LOPES simulated and measured events are pointed out.

The upper limit on the observed energy precision ($\sim$~20-25~\%) includes the convolution of both the LOPES and KASCADE-Grande (particle detector) uncertainties. The good compatibility with the KASCADE-Grande statistical uncertainty (approx.\ 20~\%) indicates that the combined uncertainty of the energy is strongly influenced by KASCADE-Grande and that the energy resolution of LOPES is likely significantly better than 20-25~\%. 
The total energy precision of the slope method suggested by the pure CoREAS simulations is approximately 13~\% and could potentially be achieved by applying the method on low-noise radio data.

It is worth stressing once again that radio emission is purely sensitive to the electromagnetic component of air showers. If combined with other detection techniques which are able to provide information on other components, in particular the muonic one, radio detection can help to disentangle the individual components, which in itself opens access to composition-sensitive information. Moreover, the radio signals are not critically affected by uncertainties in the hadronic interaction models used to understand the data.

One restriction of the analysis presented here concerns the deliberate exclusion of the detector-simulation in the investigation. A more realistic comparison with the measurements may include both simulations of detector properties as well as simulated continuous and transient noise typical of the specific experimental site.

Also, we did not try to apply a more sophisticated normalization of the radio field strengths including a charge-excess component. Since we fit a one-dimensional LDF with observers at various azimuth angles, the charge-excess asymmetry mostly averages out in the LDF fit. We confirmed with a dedicated simulation study that the additional spread introduced by this simplified treatment is negligible and that there is a bias in the reconstructed $k$-parameters of less than 5\%, which is much smaller than the overall calibration scale uncertainty of 35\% and for which we account with an additional systematic uncertainty of 5\%. For experiments with better data quality than LOPES, however, methods should be investigated to try and correct individual measured amplitudes by the presumed charge-excess contribution or fit two-dimenaional LDF functions to improve the fit and thus possibly the energy determination. This aspect was not investigated since the precision we achieve in the combined analysis anyway seems to be limited by the reference energy provided by KASCADE-Grande and not by our method.

The QGSJet~II.03 interaction model was employed to generate the CoREAS simulations of the LOPES events. Other interaction models are available and they differ both in their implementation and their predictions. We used the slope method to show how air shower parameters such as energy and X$_{\mathrm{max}}$ can be directly derived from the features of the radio LDF. Regarding the energy reconstruction, different interaction models predict different fractions of the energy going into the electromagnetic component and other components, respectively. This might lead to more significant shifts between the radio amplitudes observed for different primaries of the same energy than illustrated in this analysis. Such shifts, however, are not a disadvantage of radio detection, since in fact radio emission with its sensitivity to the pure electromagnetic component of the air shower could be used to study such effects and test hadronic interaction models. Regarding the shower evolution, different interaction models will 
make different predictions for the X$_{\mathrm{max}}$ values of air showers initiated by specific primaries with given energies. However, the obtained ``calibration'' relating a given value of X$_{\mathrm{max}}$ to a corresponding radio LDF slope should not be strongly affected by the choice of interaction model. This is because the slope method exploits an intrinsic property of the radio emission: the relation between the geometrical distance of the radio source and the slope of the radio lateral distribution arising from the forward-beaming of the emission. A slight influence on the calibration curve can be expected by variations of the atmosphere, which relate the geometrical distance to the depth of shower maximum in g/cm$^{2}$. Within the precision achievable by LOPES, however, these effects can be neglected.  
  
The X$_{\mathrm{max}}$ resolution obtained in this analysis ($\sim$95~g/cm$^{2}$) is an upper-limit attained with (LOPES) radio data affected by strong environmental noise. Much better results may be achieved by using higher quality radio measurements and a denser antenna array. Such is the case for LOFAR \cite{Buitink_icrc13}, which has recently applied a method based on global fits of the two-dimensional radio LDF and obtained a sensitivity in X$_{\mathrm{max}}$ comparable with the Fluorescence technique ($\sim$20~g/cm$^{2}$). The underlying principle is the same, namely the exploitation of the geometrical connection between the source distance and the radio LDF described above.

In the analysis presented here, we strove to exploit the potential of the LOPES data while at the same time keeping the analysis practical. Existing and upcoming more modern experiments such as AERA \cite{aera}, Tunka-Rex \cite{Tunkarex} and LOFAR \cite{Buitink_icrc13} will have higher-quality data with lower background noise, a larger number of antennas per event and a larger range of lateral distances probed with each event. The methods presented here will have to be refined to achieve the maximum precision these instruments will be able to deliver in energy and X$_{\mathrm{max}}$ resolution. In particular, a lateral distribution function accounting for the asymmetries induced by the charge-excess should be used. The function used to fit the X$_{\mathrm{max}}$ values as a function of amplitude ratios should be re-investigated and correlations between fit parameters should be minimized. Finally, a more sophisticated determination of the reconstruction uncertainties, taking into account the non-Gaussian 
error propagation should be applied. In spite of these limitations, we have demonstrated that a determination of the energy and X$_{\mathrm{max}}$ from the radio lateral distribution is feasible with promising upper limits on the achievable precision.


\section{Conclusion}

The purpose of this analysis was the reconstruction of the energy and depth of shower maximum of individual cosmic rays from radio measurements performed with the LOPES experiment. We achieved this goal by exploiting characteristic features of the lateral distribution function of the radio emission, in particular its slope.

We have developed and applied a CoREAS-calibrated method to the measurements of the LOPES experiment. Significant results emerged:

\begin{itemize}
\item[$\bullet$]
A characteristic distance from the shower axis, i.e. d$_{0}$ (70--100~m), is predicted by CoREAS simulations of the LOPES events to be the appropriate place for the primary energy investigation with radio data. Its existence is also indicated in LOPES measurements.
\item[$\bullet$] A simple linear correlation between the radio amplitude measured at the distance d$_{0}$ and the KASCADE-Grande reconstructed energy was found. A zenith-angle dependent deviation in the correlation parameters remains between CoREAS simulations and LOPES data; this discrepancy needs to be investigated in a dedicated analysis.
\item[$\bullet$] An upper-limit for the LOPES precision on the total energy reconstruction was determined from the combined LOPES--KASCADE-Grande energy uncertainty ($\sim$20-25\% depending on zenith angle). The intrinsic energy resolution of the radio measurements suggested by the simulations is approximately $13\,\%$, and might even be improved by taking the Askaryan effect explicitly into accout in the LDF fits. However, any further improvment is beyond the testing power of LOPES.
\item[$\bullet$] The slope of the LOPES lateral distribution function was exploited to reconstruct X$_{\mathrm{max}}$ from LOPES data. The resulting distributions are plausible and comparable with expectations from cosmic ray nuclei.
\item[$\bullet$] By itself, and for the specific situation of LOPES, the slope method predicted a precision of X$_{\mathrm{max}}$ around 50~g/cm$^{2}$, depending on the zenith angle of the event. From the analysis of the LOPES measurements, an upper limit of $\sim$95~g/cm$^{2}$ on
the precision was found.  
\end{itemize}

X$_{\mathrm{max}}$ is the principle indicator for cosmic ray composition, and we have demonstrated the possibility to reconstruct it with the slope method applied to LOPES radio measurements. The importance of this result is underlined by the fact that radio detection does not suffer from low duty cycles as is the case for optical detection techniques.
Comparisons of the reconstructed X$_{\mathrm{max}}$ values between, at least, two different detection methods are essential but not possible in the framework of the LOPES and KASCADE-Grande experiments. This opportunity is offered by the new generation of antenna arrays, such as AERA \cite{aera} (at the Pierre Auger Observatory) and Tunka-Rex \cite{Tunkarex}, where the radio reconstruction can be cross checked with the fluorescence and Cherenkov experimental results, respectively.

\section*{Acknowledgments}
LOPES and KASCADE-Grande have been supported by the German Federal Ministry of Education and Research. KASCADE-Grande is partly supported by the MIUR and INAF of Italy, the Polish Ministry of Science and Higher Education and by the Romanian Authority for Scientific Research UEFISCDI, IDEI, grant 271/2011. This research has been supported by grant number VH-NG-413 of the Helmholtz Association.

\end{document}